# Cavity-dumping using a microscopic Fano laser

**Gaoneng Dong,**[1,2] **Shih Lun Liang,**[1,2] **Aurimas Sakanas,**[1,2] **Elizaveta Semenova,**[1,2] **Kresten Yvind,**[1,2] **Jesper Mørk,**[1,2,*] **And Yi Yu**[1,2,*]

[1]*DTU Electro, Technical University of Denmark, Lyngby, Denmark.*

[2]*NanoPhoton – Center for Nanophotonics, Technical University of Denmark, Lyngby, Denmark.*

*jesm@fotonik.dtu.dk, yiyu@fotonik.dtu.dk*

**Abstract:** A microlaser with low energy consumption and high speed is crucial for on-chip photonic networks. Presently, the modulation of microlasers is based on modulating the gain of the laser, which implies a trade-off between the output peak power and the modulation energy. Besides, the temporal width of the output pulse is restricted by the carrier relaxation time. These limitations can be overcome by modulating, instead, the loss of the laser by the scheme of cavity dumping, which is ideal for intense and ultrashort pulse extraction. However, the miniaturization of cavity-dumped lasers has been a long-standing challenge, and no microscopic cavity-dumped lasers were yet realized. Here we demonstrate an ultra-small cavity-dumped microscopic laser based on an optical Fano resonance, which generates optical pulses with peak power more than one order of magnitude higher than the corresponding conventional gain-modulated laser. This demonstration paves the way for realizing microscopic lasers for low-power chip-scale applications.

# 1. Introduction

On-chip optical interconnects [1] have the potential to overcome the limits on bandwidth and power consumption in traditional electrical interconnects [2]. The transmitter in an on-chip optical network needs to meet several demanding requirements, i.e., sufficient peak power, high speed, low energy consumption, and ultra-small footprint. One well-known transmitter configuration integrates external modulators [3, 4] with continuous-wave (CW) microlasers [5, 6]. Another simpler one is the directly modulated microlaser. The latter generally has much lower power consumption, smaller size, and lower cost, and has been widely investigated [7-14]. However, these microlasers are all based on modulating the gain of the laser, implying a severe trade-off between the output peak power and the modulation energy [14]. In addition, the modulation bandwidth is inherently limited by the relaxation oscillation frequency.

One way to overcome these obstacles is to use the idea of cavity-dumping, where the output power is modulated via the cavity loss rather than the gain of the laser. The cavity-dumped laser was first demonstrated more than 50 years ago [15], and has been applied in numerous schemes to generate high-peak-power pulses with pulse durations from nanosecond [16-19] to femtosecond (combining with mode-locked technique) [20-23]. However, these schemes are traditionally based on large and complicated multi-element systems with meter-sized cavity lengths. Recently, several integrated lasers exploiting reflectivity-modulation were demonstrated [24-27]. Tessler et al. reported a wide-band amplitude modulation edge-emitting laser [27]; Paraskevopoulos et al. demonstrated a large-bandwidth vertical-cavity surface-emitting laser (VCSEL) [26], both through tuning the stopband edge of distributed Bragg reflectors (DBRs); Dong et al. reported directly reflectivity-modulated lasers by employing a composite mirror [24, 25]. These systems, however, are still macroscopic with relatively large footprint, which are complex and energy consuming.

Here we demonstrate a microscopic cavity-dumped laser with a mirror based on optical Fano resonance [28, 29]. The rich physics of Fano resonances has been studied in numerous photonic and plasmonic nanostructures [30, 31]. In particular, by replacing one laser mirror with a narrowband mirror, the so-called Fano mirror, a Fano laser has been achieved [32, 33], showing several important features, including suppression of feedback-induced instabilities [34], self-pulsing [32], and orders-of-magnitude reduction of the laser linewidth [35]. In this work, we experimentally and theoretically demonstrate that the output power of a Fano laser can be efficiently modulated via the nanocavity to perform cavity dumping, which generates optical pulses with peak power more than one order of magnitude higher than that of an equivalent gain-modulated laser.

## 2. Concept, schematic, and principle

The conventional way of modulating a laser is to modulate the gain, cf. Fig. 1(a) (here, we refer to the laser bias as the pumping source). Such a modulation scheme typically cannot generate high-peak-power pulses since the maximum output power of the laser is limited by the modulation depth $\Delta R/R$, e.g., if the initial constant bias is above the laser threshold (called bias-modulation), the peak modulated power scales as $P_{b,peak} = P_s + \eta_p (\Delta R/R) P_s$, where $P_s$ is the initial output power at steady state, which is proportional to the initial pumping rate, $R$, $\Delta R$ is the change of the pumping rate due to the modulation signal and $\eta_p$ is a coefficient smaller than one (Supplement Note A.1). In contrast, the cavity-dumped laser can modulate the output signal by modulating one of its mirrors, which acts as a switchable "gate". For example, by applying an external optical pulse, the right mirror "opens" (Fig. 1(b)), and a large number of photons initially stored in the laser cavity can thus be dumped within a short time, generating a high-peak-power optical pulse at the output. In this case, the maximum output power of the pulse is (Supplement Note A.1)

$$P_{\text{d,peak}} = P_{\text{s}} + \frac{\Delta r_{\text{R}} \left(2r_{\text{R}} - \Delta r_{\text{R}}\right)}{t_{\text{R}}^2} P_{\text{s}}, \tag{1}$$

where $r_{\text{R}}$ ($t_{\text{R}}$) is the right mirror reflectivity (transmissivity) at steady state, and $\Delta r_{\text{R}}$ is the amplitude of the maximum reflectivity change. Since usually $r_{\text{R}} \rightarrow 1$ and $t_{\text{R}}^2 \ll 1$, $P_{\text{d,peak}}$ can be much larger than $P_{\text{b,peak}}$ (Supplement Note A.1). It should be noted that by modulating the gain when the initial constant bias is below the laser threshold (called gain-switching [36]), one can also generate optical pulses, similar to the bias-modulation scheme, however, the maximum power $P_{\text{g,peak}}$ is still limited by the pumping rate, and much lower than $P_{\text{d,peak}}$ (Supplement Note A.1). Additionally, the time duration of the cavity-dumped output pulse, unless limited by the switching time of the laser mirror, will be determined by the roundtrip time of the laser cavity. This roundtrip time can be very short (femtoseconds) for microscopic lasers (Supplement Note A.1); orders of magnitude faster than the conventional gain-modulated scheme where the pulse duration is limited by the carrier relaxation time (typically on the order of several picoseconds even for high pumping rates).

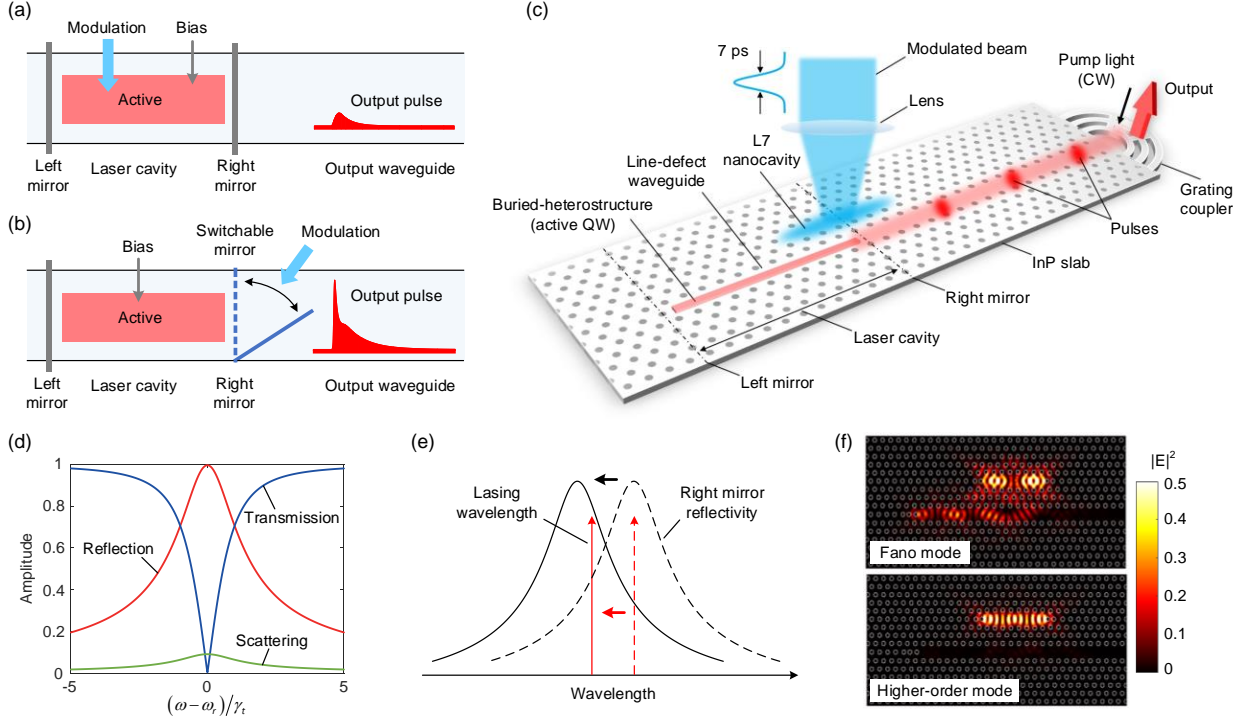

**Fig. 1.** Concept, schematic, and principle of cavity-dumped Fano laser. (a) Concept of a conventional gain-modulated laser. An external modulation changes the gain of the active region in the laser cavity, resulting in the generation of an optical pulse. (b) Concept of a cavity-dumped laser. The reflectivity of one of its mirrors is changed by the external modulation, releasing the optical energy in the laser cavity within a short time and thereby generating an optical pulse with high peak power in the output waveguide. (c) Schematic of the optically modulated cavity-dumped Fano laser based on a Fano mirror formed by coupling a photonic-crystal waveguide with a side-coupled nanocavity. (d) Example of calculated reflectivity (red line), transmissivity (blue line), and out-of-plane scattering (green line) of a Fano mirror. The intrinsic Q-factor $Q_v$ and coupling Q-factor $Q_c$ of the nanocavity are 120,000 and 500, respectively. (e) Principle of the switchable (Fano) mirror. With the blue shift of the nanocavity resonance, the lasing wavelength (solid vertical red arrow) will partially track the peak of the reflection spectrum (solid black curve), but the effective detuning will increase, leading to a reduction of the reflectivity. (f) Calculated (normalized) optical intensity profiles ($|E|^2$) of the two relevant optical modes, both shown in the center plane of the PhC membrane. The colorbar is saturated at 0.5 for a better view. The upper panel shows the Fano mode (corresponding to the second-order mode of the L7 nanocavity), in which the laser oscillates, and the lower panel shows a higher-order mode (corresponding to the third-order mode of the L7 nanocavity), which is mainly concentrated in the L7 nanocavity and used for modulating the nanocavity and thereby the right mirror.

Here, we realize a microscopic cavity-dumped laser that exploits optical Fano resonance. The laser is based on a photonic crystal (PhC) membrane structure on silicon composed of a line-defect waveguide (WG) and two qualitatively different mirrors (Fig. 1(c)), with an effective footprint of only 12.5 × 6 μm². The left mirror is a conventional broadband mirror formed by simply terminating the WG with holes, while the right switchable mirror is based on a Fano resonance that originates from the interaction between a continuum of WG modes with a discrete mode of a side-coupled nanocavity. Specifically, the discrete mode is the second-order mode of a L7 nanocavity formed by omitting seven air holes. Due to the Fano destructive interference at the output of the right mirror, wavelengths close to the resonance of the nanocavity will be reflected with a near-unity reflection coefficient (Fig. 1(d)). If the total nanocavity decay rate is dominated by the coupling rate $\gamma_c$ between the nanocavity and the WG, rather than by the intrinsic decay rate $\gamma_v$ of the nanocavity, the transmissivity ($\gamma_v/(\gamma_v + \gamma_c)$) at the nanocavity resonance will be much smaller than the out-of-plane scattering of the nanocavity ($\sqrt{\gamma_c \gamma_v}/(\gamma_v + \gamma_c)$). This feature of the Fano mirror is favorable for generating optical pulses with high contrast in the output waveguide (through-port). Similar to our recent demonstration of an ultra-coherent Fano laser [35], we use a buried-heterostructure (BH) region [12] containing a single InGaAsP/InAlGaAs quantum well (Fig. 2(a)) to confine the active material to the laser cavity while leaving the rest area passive. Here, we refer to the laser cavity as the region between the left mirror and the Fano mirror in order to separate it from the nanocavity. Details of the design and fabrication can be found in Ref. [35].

The Fano laser is optically pumped by a CW laser via a broadband grating coupler [37] located at the right end of the WG (Fig. 1(c)). The detailed modulation principle is illustrated in Fig. 1(e). At steady state, the lasing wavelength is aligned to the peak reflectivity of the narrowband Fano mirror so the optical field is well confined in the laser cavity and the nanocavity (see the upper panel of Fig. 1(f)). An external optical pulse is injected from the top and is coupled into the third-order mode of the nanocavity. This mode, in contrast to the second-order mode of the nanocavity, is highly localized at the nanocavity, i.e., it hardly couples to the WG mode (see the bottom panel of Fig. 1(f)). Such a strong field localization not only causes an effective local index change through optical nonlinearities that quickly blue shifts the resonant frequency of the nanocavity, but also ensure that the nanocavity rather than the entire laser gets modulated. During the resonance shift, the lasing wavelength will partially track the nanocavity resonance, but with an increasing effective detuning [33], leading to a reduction (increase) of the reflectivity (transmissivity) of the laser mirror (Fig. 1(e)). Thus, the photons stored in the laser cavity is discharged from the through-port.

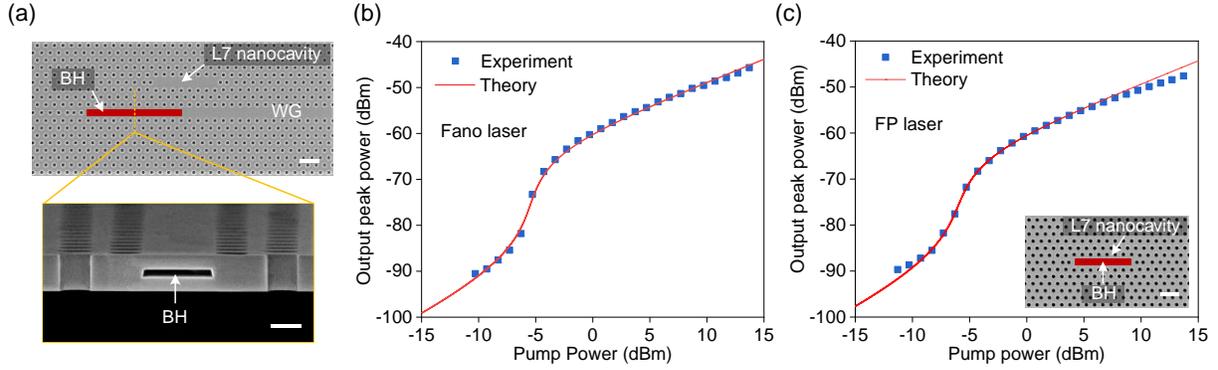

**Fig. 2.** Device and statistic characteristics. (a) Scanning electron microscope (SEM) image of a fabricated Fano laser (upper panel) based on an InP PhC membrane structure embedded with a buried heterostructure (BH, red rectangle) gain region composed of a single quantum well. The side-coupled L7 nanocavity is passive. Scale bar: 1 μm. The inset (lower panel) is the SEM image of the cross-section of a fabricated Fano laser showing the active waveguide containing a BH. Scale bar: 200 nm. (b) – (c) Collected output peak power of the laser spectra (OSA resolution: 0.1 nm) versus pump power for the Fano laser (b) and the FP laser (c). The pump power refers to the value after the microscope's objective lens. The blue squares are experimental data, and the red lines are theoretical results based on rate equations with fitted parameters. The inset in (c) shows the SEM image of the FP laser with a BH (red rectangle) embedded only within the L7 nanocavity. Scale bar: 1 μm.

## 3. Demonstration of cavity dumping

The Fano laser oscillates in a single mode, as observed in our previous works [32, 35]. Figure 2(b) depicts the output peak power versus pump power, showing a clear transition to lasing at a threshold pump power of -5 dBm. The measured maximum output power is about -50 dBm. For comparison, a PhC line-defect laser, which can effectively be considered as a Fabry–Pérot (FP) laser, is also characterized (Fig. 2(c)). The FP laser is identical to the side-coupled nanocavity used in the Fano laser except for embedding an active BH into the nanocavity. The FP laser is vertically pumped with the same light source as the Fano laser, and the emission is collected vertically (Supplement Note B.1). Although the pumping and collection schemes differ, the two lasers show very similar characteristics in terms of laser threshold and collected output power.

To verify the observation of cavity dumping, we measure the dynamics of the output signal of the Fano laser coupled via the through-port (emission from the grating coupler) and the cross-port (emission from the nanocavity) (Supplement Note B.2). We also characterize the dynamics of the FP laser by aligning the modulating pulse to the third-order mode of the L7 nanocavity (Supplement Note B.2), where the pump and modulation powers are the same as for the Fano laser.

The output waveforms (Fig. 3(a), after subtracting the noise of the Erbium-doped fiber amplifier (EDFA) (Supplement Note D)) of the Fano laser differ qualitatively between the through- and cross-port. The through-port signal exhibits a high-contrast "peak", while a "dip" appears in the cross-port signal (Fig. 3(a)). The cross-port signal directly reflects the energy stored in the laser cavity, which follows an opposite time evolution as the through-port signal that reflects the cavity-dumped energy. This is a typical feature of cavity-dumped lasers [16, 18, 19]. In contrast, the output signal from the FP laser evolves in the same way as the injected signal, reflecting the conventional gain-modulation scheme.

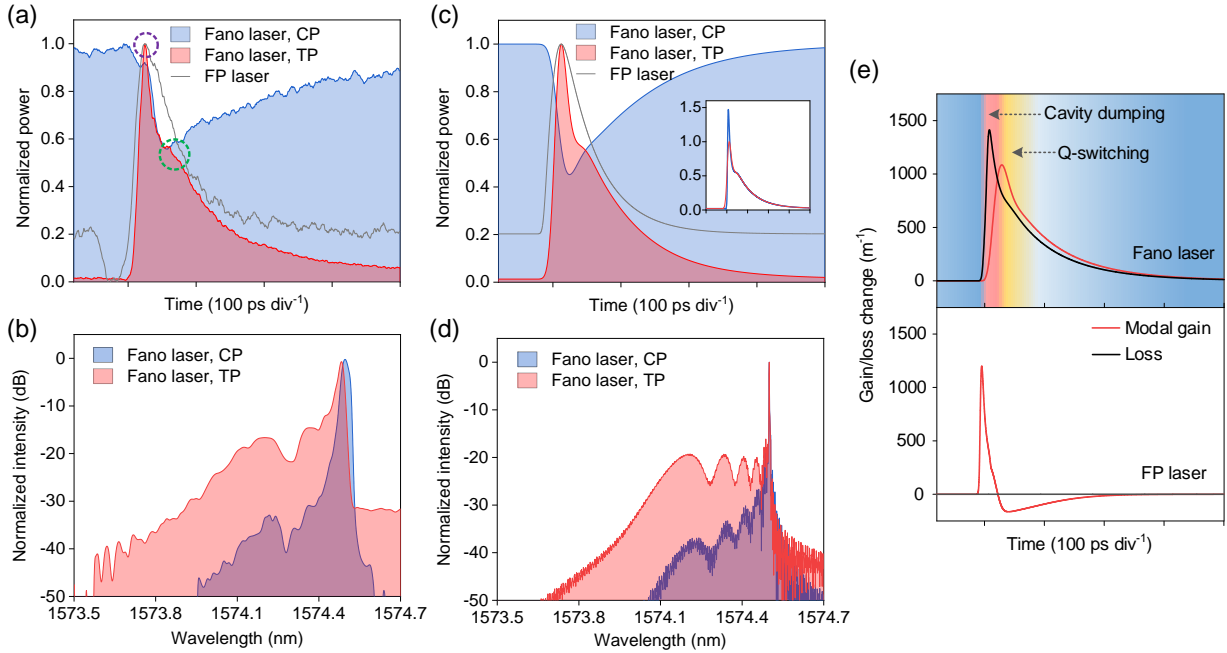

**Fig. 3.** Demonstration of cavity-dumping in a Fano laser. (a) – (b) Measured (normalized) waveforms (a) and (normalized) spectra (b) of the through-port (TP) and cross-port (CP) signals of the Fano laser. The main peak of the TP signal in (a) (marked by the purple dashed circle) is generated by cavity dumping, while the following kink (marked by the green dashed circle) is due to cavity Q-switching. The grey curve in (a) represents the measured (normalized) waveform of the FP laser. The pumping rate and average modulation power are 79 $P_{th}$ and 1.26 mW, respectively for both lasers. (c) – (d) Calculated (normalized) waveforms (c) and (normalized) spectra (d) of the TP and CP signals of the cavity-dumped Fano laser. The grey curve in (c) represents the calculated (normalized) waveform of the FP laser. The pumping rate and average modulation power are 79 $P_{th}$ and 1.26 mW, respectively for both lasers. Note that the calculated waveforms have been convoluted with the oscilloscope's pulse response (16.5 ps). Inset in (c): the calculated waveforms from the TP of the Fano laser with (red curve) and without (blue curve) convolution with the oscilloscope's pulse response. They are both normalized by the peak of the red curve. (e) Calculated time dependence of the gain (red curve) and loss (black curve) change for the Fano laser (upper panel) and the FP laser (lower panel) during the modulation. In the upper panel, the pink (yellow) shaded area refers to the cavity dumping (Q-switching) region of the Fano laser.

The measured spectra of the Fano laser are asymmetric with an enhanced blue component in both the through- and cross-port signals (Fig. 3(b)), which is in accordance with the modulation mechanism of the Fano mirror (Fig. 1(e)). The spectral broadening is much larger for the through-port signal than for the cross-port signal, which is consistent with their waveforms (Fig. 3(a)). The measurements can be well fitted by our numerical simulations (Supplement Note A.2, A.3 and C.2), cf. Fig. 3(c) and (d), showing that the loss of the Fano laser is varying in time and is governed by the mirror loss, which increases rapidly after the arrival of the modulating pulse (see the upper plane of Fig. 3(e)) due to the resonance shift of the nanocavity. The Fano laser transitions from the initial "ON" state to the "OFF" state as the loss suddenly exceeds the gain. In contrast, the loss of the FP laser remains constant (see the lower plane of Fig. 3(e)), in accordance with the gain modulation due to linear absorption of the modulated light in the third-order mode of the L7 nanocavity (Supplement Note C.3). Although the nonlinear absorption in the nanocavity (including two-photon absorption and free-carrier absorption) can also lead to the transmissivity increase of the Fano mirror, it is unfavorable for high-power pulse generation, causing a reduction of the peak power by 30% according to simulations. It should be emphasized that for the Fano laser, the measured waveforms are limited by the time resolution of the oscilloscope, which is not the case for the FP laser due to the long duration of its output pulse. The actual pulse peak power of the Fano laser is 50% higher than the measured value (see the inset in Fig. 3(c)). Although cavity dumping is an effective approach to generate ultrashort pulses, in our case, the pulse suffers from a relatively long tail which is due to the slow lifetime of free carriers generated in the nanocavity [38]. In addition, following the cavity dumping, when the gain eventually exceeds the loss as the Fano resonance shifts back (see the upper panel of Fig. 3(e)), the laser transitions from the "OFF" state to the "ON" state, releasing a "secondary pulse" due to cavity Q-switching [33], thus adding a kink and elongating the tail of the output waveform, cf. Fig. 3(a) and (c). Our theoretical model predicts that with a larger modulation power, the "primary" and the "secondary" pulses can be well separated in time, which narrows the time width of the main pulse (Supplement Note C.2). This indicates that a short pulse may still be generated even with the presence of a slow carrier lifetime.

## 4. Dynamic characteristics

Next, we compare the dynamics of the Fano laser and the FP laser in dependence of the modulation and pump powers. The power value is measured at the final detector, and the estimated ratio of the coupling efficiency (from the output of the laser mirror to the detector) is about 2 ~ 6 for the Fano laser with respect to the FP laser. The peak power of the Fano laser increases efficiently with both the modulation power (Fig. 4(a)) and the pump power (Fig. 4(d)), which is not the case for the FP laser (Fig. 4(b) and (e)). This is also reflected in their corresponding spectra (Supplement Note C.1), and in accordance with our simulations (Supplement Note C.2). A clearer comparison is illustrated in Fig. 4(c) (varying the average modulation power while keeping the pump power

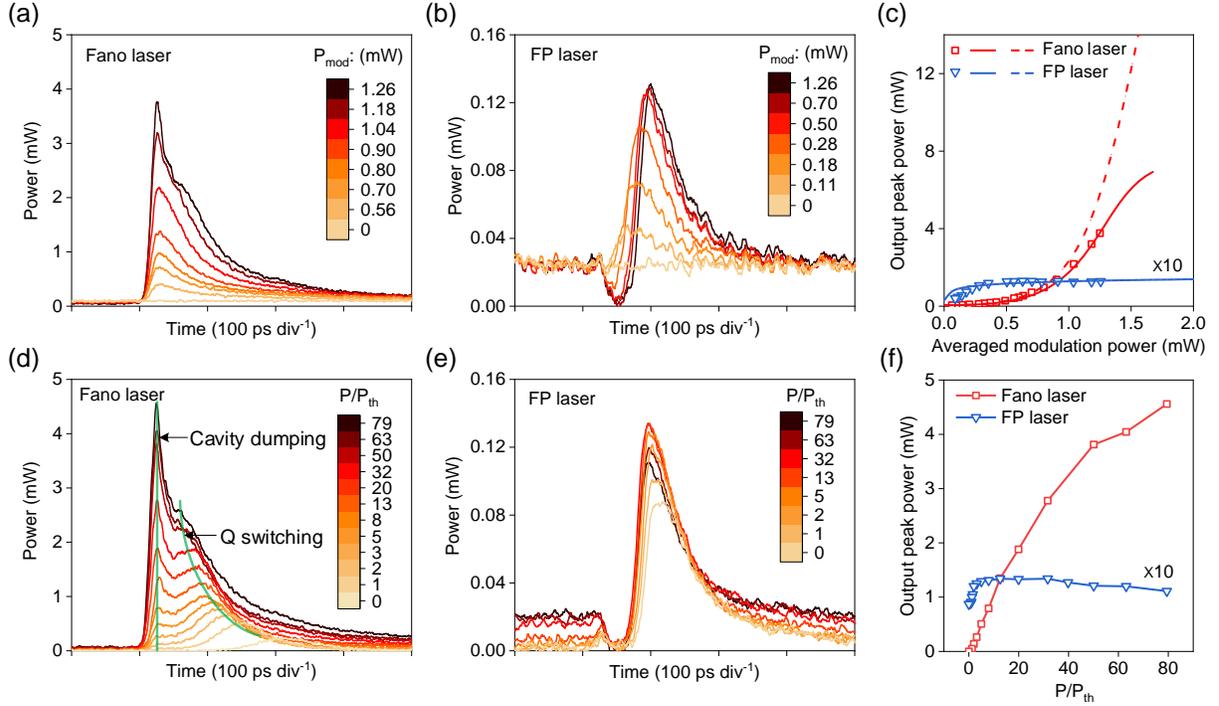

**Fig. 4.** Dynamic characteristics of cavity-dumped Fano laser. (a) – (b) Measured waveforms of the through-port signal of the Fano laser (a) and the signal emitted from the top of the FP laser (b) for different values of modulation power. The pumping rate is $63\,P_{th}$ for both lasers. The average modulation power refers to the value measured after the microscope's objective lens. (c) Output peak power versus average modulation power. The symbols are the measured results, while the solid (dashed) line is the calculated results after (before) the convolution with the oscilloscope's pulse response. The solid and dashed lines overlap in the FP laser. The calculated output peak power is amplified by 28 dB for both lasers to match the experimental results (with the EDFA amplification in front of the oscilloscope). (d) – (e) Measured waveforms of the through-port signal of the Fano laser (d) and the signal emitted from the top of the FP laser (e) with varied pump power. The average modulation power is fixed at 1.26 mW for both lasers. In (d), the left green line indicates the peak associated with cavity dumping and the right green curve indicates the peak associated with cavity Q-switching. (f) Output peak power (refers to the narrow front peak in (d)) versus pump power. In (c) and (f), the output peak power of the FP laser is multiplied by another factor of 10 to make it clearer for comparison.

constant) and Fig. 4(f) (varying the pump power while keeping the modulation power constant).

The peak power of the output pulse from the Fano laser is more than one order of magnitude higher than the FP laser (11 ~ 33 times when normalized by the coupling efficiency). According to our simulations, a further improvement by more than two orders of magnitude should be achievable for higher modulation powers (Fig. 4(c)), which is, however, not attainable in our experimental setup. This reflects the intrinsic advantage of the cavity-dumping scheme enabled by the Fano laser compared to the gain-modulation scheme, cf. earlier discussion of the trade-off between the output peak power and the modulating pulse energy (Supplement Note A.1 and A.3). For optical pumping, the ratio of the output peak power with respect to the average input modulation power, $P_\text{o}^\text{p}/P_\text{i}^\text{p}$, increases monotonically with $P_\text{i}^\text{p}$ for the Fano laser (Fig. 4(c), Supplement Note A.2), while the peak power of the FP laser saturates (Fig. 4(c)), which is due to carrier density saturation under optical modulation (Supplement Note A.3). Although such a saturation effect is expected to be eliminated under electrical modulation, which is beyond the scope of this work and requires further investigations, our preliminary simulations show that compared to the FP laser, the pulse peak power of the Fano laser is still much higher and increases much faster with the modulation/pump power, due to the nature of cavity-dumping.

## 5. Discussion and conclusion

Theoretical fits to the experimental results suggest that the initial lasing wavelength of the Fano laser is slightly detuned with respect to the peak of the Fano mirror reflectivity due to fabrication imperfections, which degrades the quality of the output pulse. This can be alleviated by placing microheaters [39] in the surroundings to tune the nanocavity resonance or the phase delay in the WG. Besides, the slow carrier lifetime in the nanocavity sets a limit on the pulse duration and thus the modulation speed of the Fano laser. This problem might be overcome by employing carrier sweeping [40], or replacing the nanocavity material with one where instantaneous optical Kerr [41], or Pockels electro-optic [42] effects dominate. The frequency chirping imposed on the out-coupled pulse, appearing to be unavoidable for the Fano laser modulation, is not a big problem for on-chip applications because the optical interconnect distances are so short that the effects of the pulse dispersion become negligible. According to our simulations, although the absorbed energy for modulation (~ 16 fJ/pulse) is currently larger than the generated pulse energy (~ 0.85 fJ/pulse), the modulation energy can be significantly decreased by reducing the mode volume of the nanocavity, which can lead to a higher localization of the nanocavity field and thus stronger optical nonlinearities, as verified by PhC nanocavity switches [38, 43]. For example, if the mode volume of the nanocavity is reduced by a factor of 20, which is feasible by employing an extremely dielectric cavity [44], the generated pulse energy (0.85 fJ/pulse) will exceed the modulation energy (0.83 fJ/pulse), acting as an optical transistor. Besides, a more efficient modulation associated with a smaller mode volume could also help narrow the pulse width (Supplement Note C.2).

In summary, we have experimentally demonstrated an ultra-small cavity-dumped microscopic laser based on optical Fano resonance. By optically modulating the nanocavity-based Fano mirror, the laser generates optical pulses with peak power more than one order of magnitude higher than a corresponding gain-modulated FP laser. The cavity-dumping scheme relaxes the trade-off between the output peak power and the modulating energy, an inherent issue for conventional gain-modulated lasers, and the small mode volume of the nanocavity helps reduce the energy consumption. Our scheme based on optical Fano resonance can be extended to other configurations, including vertical [6] and hybrid integration [6, 45], and can be combined with electrical modulation [46, 47], which is essential for low-power, ultrafast chip-scale applications including conventional optical communication and computation [48-50], as well as spiking neuromorphic networks [51-53].

**Funding.** European Research Council (ERC) under the European Union Horizon 2020 Research and Innovation Programme (Grant no. 834410 FANO); Danish National Research Foundation (Grant no. DNRF147 NanoPhoton); Villum Fonden via the NATEC Center (Grant no. 8692); Villum Fonden via the Young Investigator Program (Grant no. 42026).

**Acknowledgments.** We thank M. Galili, M.H. Pu, and L.K. Oxenløwe for assistance with experimental equipment, and M. Xiong for inductively coupled plasma etching optimization and assistance with sample characterization.

**Disclosures.** The authors declare no competing interests.

**Data availability.** Data underlying the results presented in this paper may be obtained from the authors upon reasonable request.

**Supplemental document.** See Supplement 1 for supporting content.

**Supplemental Document: Cavity-dumping using a microscopic Fano laser**

Gaoneng Dong, Shih Lun Liang, Aurimas Sakanas, Elizaveta Semenova, Kresten Yvind, Jesper Mørk[*], and Yi Yu[*]

*DTU Electro, Technical University of Denmark, Lyngby, Denmark.*
*NanoPhoton – Center for Nanophotonics, Technical University of Denmark, Lyngby, Denmark.*
*jesm@fotonik.dtu.dk, yiyu@fotonik.dtu.dk*

## A. Theory

### A.1. Cavity dumping

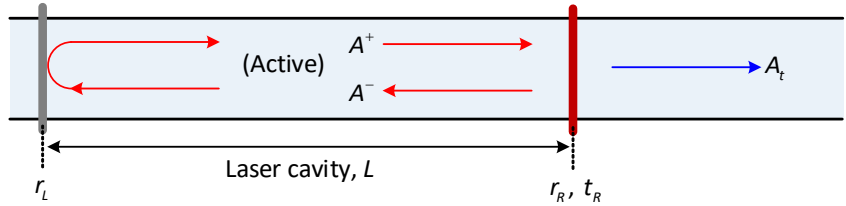

**Fig. S1**. Schematic of a Fabry–Pérot laser whose right mirror reflectivity is variable.

Here we explain the cavity dumping by a simple toy model, which agrees very well with the more realistic model described in section A.2. For a Fabry–Pérot (FP) laser (Fig. S1) having a lossless broadband left mirror ($r_L = 1$) and a right mirror whose reflectivity can be changed (like the Fano laser), the laser field (assumed to be real) and carrier density obey [1, 2]

$$\frac{dA^+(t)}{dt} = \left(\frac{1}{2}G_N(N(t)-N_{tr})-(\gamma_{in}+\gamma_p/2)\right)A^+(t) + \frac{1}{2}\left(\Gamma\beta_{sp}\frac{N(t)}{\tau_s}\right)\frac{V_p}{I(t)}A^+(t) + \frac{\gamma_{in}A^-(t)}{r_{R,s}}, \quad (s1)$$

$$\frac{dN(t)}{dt} = R - \frac{N(t)}{\tau_s} - v_g g_N(N(t)-N_{tr})\frac{I(t)}{V_p}, \quad (s2)$$

where $A^+(t)$ ($A^-(t) = r_R(t)A^+(t)$) is the forward (backward) propagating wave to the right mirror, with $r_R(t)$ being the real-time reflectivity (assumed to be also real), $G_N = \Gamma v_g g_N$ is the optical gain coefficient, with $\Gamma$ being the optical confinement factor, $v_g = c/n$ being the material group velocity, $g_N$ being the differential gain of the active material. $N(t)$ is the free carrier density, $N_{tr}$ is the carrier density at transparency, $\gamma_{in} = v_g/2L$ is the inverse round-trip time of the laser cavity, with $L$ being the laser cavity length. $\gamma_p$ is the inverse of photon lifetime $\tau_p$, $\beta_{sp}$ is the spontaneous emission factor, $\tau_s$ is the carrier lifetime, $V_P = A_0 L$ is the optical mode volume of the laser cavity, with $A_0$ being the area of the cross-section of the optical mode in the waveguide, $r_{R,s}$ is the right mirror reflectivity at steady state, $R$ is the pumping rate, and $I(t) = \sigma_s |A^+(t)|^2$ is the number of photons stored in the laser cavity, with the factor $\sigma_s$ calculated at steady state, cf. (s23) in section A.2. The cavity dumping is initiated by opening the right mirror (decreasing the mirror reflectivity). Here, we employ a simple model for the change of $r_R(t)$,

$$r_R(t) = r_{R,s} - \Delta r_R \left( \Theta(t-t_0) \exp(-\gamma_m (t-t_0)) + \Theta(t_0-t) \exp(-(t-t_0)^2/(2t_1^2)) \right), \quad (s3)$$

where $\Delta r_R$, $t_0$ are the amplitude and time point of the maximum reflectivity change, $\gamma_m$ is the characteristic decay rate, $t_1$ is the characteristic rising time constant, and $\Theta(t)$ is the Heaviside step function. The output power through the right mirror is

$$P_o(t) = |A_t(t)|^2 = |t_R(t)|^2 |A^+(t)|^2, \quad (s4)$$

where $t_R(t) = \sqrt{1 - r_R(t)^2}$ is the real-time transmissivity of the right mirror. Next, we focus on two key features of the output pulse: the pulse decay rate and the peak power.

By neglecting the spontaneous emission term, time integrating the right part of Eq. (s1), and focusing on the trailing part of the output pulse ($t \geq t_0$), we find
$$P_o(t) \propto$$
$$\left( t_{R,s}^2 + \Delta r_R \exp(-\gamma_m (t-t_0)) \right) \left( 2 r_{R,s} - \Delta r_R \exp(-\gamma_m (t-t_0)) \right) \exp\left( -\frac{2 \Delta r_R \gamma_{in}}{r_{R,s} \gamma_m} \left( 1 - \exp(-\gamma_m (t-t_0)) \right) \right),$$
(s5)

where $t_{R,s}$ is the right mirror transmissivity at steady state. Numerical simulations of Eq. (s5) show that for a strong perturbation ($\Delta r_R \to r_{R,s}$), the decay rate of the output pulse is determined by the larger value of $\gamma_m$ and $\gamma_{in}$. Therefore, femtosecond pulse can be expected in microscopic lasers. This is orders of magnitude shorter than for conventional modulation schemes, such as gain modulation, where the pulse decay rate is limited either by the carrier lifetime (low pumping case, on the order of ns) or the stimulated emission (high pumping case, on the order of ps).

The peak power of the output pulse, found from Eqs. (s1) - (s4), is

$$P_{d,peak} = P_s + \frac{\Delta r_R (2 r_{R,s} - \Delta r_R)}{t_{R,s}^2} P_s, \quad (s6)$$

where $P_s$ is the initial output power at steady state. In contrast, in the case of gain modulation, which can be divided into two categories: bias modulation (the bias is above threshold) and gain switching (the bias is zero or below threshold), the peak power of the bias-modulation and gain-switching schemes are limited by

$$P_{b,peak} = P_s + \eta_p \left(\Delta R/R\right) P_s, \tag{s7}$$

and

$$P_{g,peak} = f\left(\Delta R, \beta_{sp}\right) \eta_p \left(\Delta R/R\right) P_s, \tag{s8}$$

respectively. Here $\Delta R$ is the amplitude change of the pumping rate, and $\eta_p$ is a coefficient usually much smaller than one for fast modulation, as limited by the relatively long carrier relaxation time. Like $\eta_p$, the factor $f\left(\Delta R, \beta_{sp}\right)$ in Eq. (s8) lacks an analytical form but is proportional to $\Delta R$ and inversely proportional to $\beta_{sp}$ [3]. For microscopic lasers which in general have a relatively large $\beta_{sp}$, $f\left(\Delta R, \beta_{sp}\right)$ is usually smaller than 3. It should be emphasized that $P_{g,peak}$ is independent of $R$, because $P_s$ is proportional to $R$. The power $P_{g,peak}$ is normalized by $P_s$ for a more straightforward comparison. Since ordinarily $r_{R,s} \to 1$ and $t_{R,s}^2 \ll 1$, by comparing Eq. (s6) to Eqs. (s7) and (s8), we can see that the power variation will be much larger to modulate $\Delta r_R$ than $\Delta R$, especially for fast modulation. Thus, compared to the ordinary gain modulation scheme, cavity dumping can be used for the generation of shorter pulses with higher peak power.

Fig. S2 shows example comparisons of the lasing dynamics for different modulation schemes for a typical microscopic laser, whose cavity length and intrinsic quality factor are 5.5 μm and 10000 (corresponding to a right mirror reflectivity of 0.993), respectively. The laser is pumped by a step function set at $t = 0$, then modulated by a pump pulse at $t = 1$ ns, followed by a right mirror reflectivity modulation at $t = 2$ ns. As seen from Fig. S2 (a), although the gain switching is able to generate a higher peak-power pulse than the bias modulation, its peak power drops severely as $\beta_{sp}$ increases from 0.001 to 0.1 (Fig. S2 (b)) as the laser enters the microscopic regime [4], [5]. In any case, the power variation of the cavity-dumping scheme is much larger than the gain modulation scheme even for a small reflectivity change, which is consistent with the above analytical analysis.

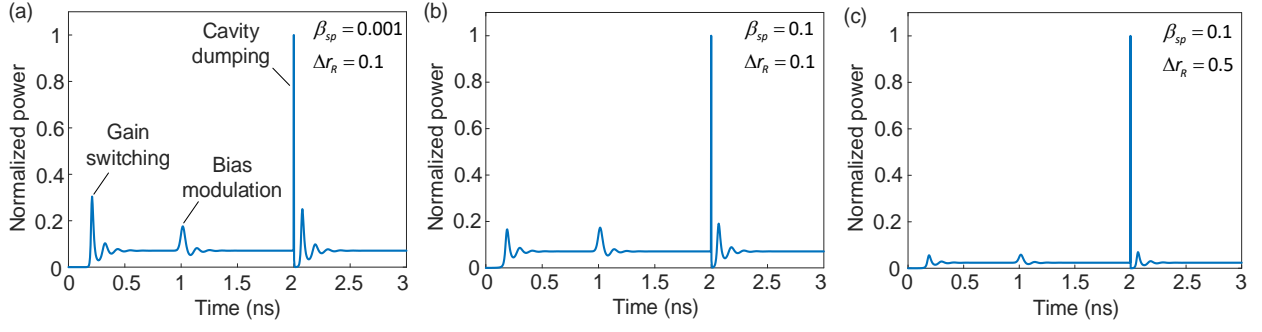

**Fig. S2**. Comparisons of output power dynamics for cavity dumping, bias modulation, and gain switching for different spontaneous emission factors $\beta_{sp}$ and maximum reflectivity change $\Delta r_R$. (a) $\beta_{sp}=0.001$, $\Delta r_R=0.1$. (b) $\beta_{sp}=0.1$, $\Delta r_R=0.1$. (c) $\beta_{sp}=0.1$, $\Delta r_R=0.5$. The other parameters are: $L=5.5$ μm, $r_{R,s}=0.993$, $R=\Delta R=10R_{th}$. The pulse width of the modulated pulse of bias modulation is set to 50 ps, so that $\eta_p \approx 1$.

## A.2. Model of the cavity-dumped Fano laser

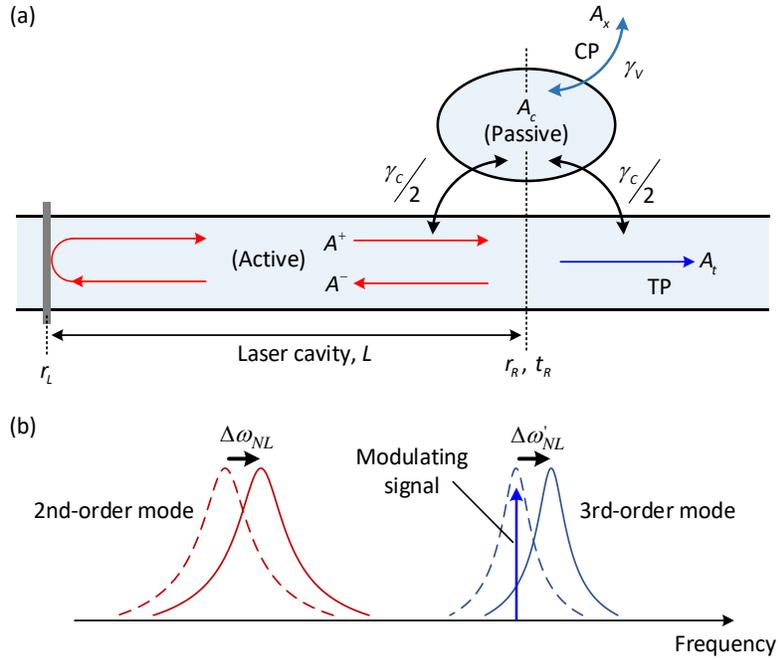

**Fig. S3**. (a) Schematic of the cavity-dumped Fano laser. (b) Modulation principle of the Fano mirror.

The dynamics of the Fano laser are efficiently modeled by an iterative equation for the field [6]

$$A^+(t+\tau_{in}) = r_L \exp\left[2iLk_s(\omega_s, N_s)\right]\exp\left(\frac{1}{2}(1-i\alpha)G_N(N(t)-N_s)\tau_{in}\right)A^-(t), \quad (s9)$$

and a conventional rate equation for the carrier density as Eq. (s2). Here $r_L$ is the reflectivity of the left mirror, which is a constant. $k_s(\omega_s, N_s)$ is the complex wavenumber at steady state, with $\omega_s$ being the lasing frequency, and $N_s$ being the carrier density in the laser cavity. The second exponential term at the right side of Eq. (s9) represents the phase and gain perturbation, with $\alpha$ and $\tau_{in} = 2\sigma L/v_g$ being the Henry's factor, and the round-trip time of the laser cavity, respectively. $\sigma$ is the slow-light factor of the photonic crystal (PhC) line-defect waveguide (WG). For the lasing field in the nanocavity $A_c(t)$, we have

$$\frac{dA_c(t)}{dt} = (-i\delta_c - \gamma_t)A_c(t) + \sqrt{\gamma_c}A^+(t)$$
$$A^-(t) = \sqrt{\gamma_c}A_c(t)$$
$$A_t(t) = A^+(t) - \sqrt{\gamma_c}A_c(t) \qquad (s10)$$
$$A_x(t) = \sqrt{\gamma_v}A_c(t),$$

where $\gamma_t = \gamma_c + \gamma_v$ is the total decay rate of the nanocavity, with $\gamma_c$ accounting for the coupling between the nanocavity and the WG, and $\gamma_v$ accounting for the vertical emission. They are related to the corresponding Q-factors of the nanocavity by $\gamma_x = \omega_0/(2Q_x)$, $(x = v, c, t)$, where $Q_v$, $Q_c$ and $Q_t$ are the intrinsic, coupling, and total Q-factors of the nanocavity, respectively. $\delta_c = \omega_0 + \Delta\omega_{NL} - \omega$ is the complex frequency detuning between the nanocavity resonance $\omega_0 + \Delta\omega_{NL}$ and the frequency of the right-propagating wave $\omega$, where $\omega_0$ is the unperturbed nanocavity resonance frequency and $\Delta\omega_{NL}$ is the complex frequency change of nanocavity. The output power in the through-port (TP) and cross-port (CP) of the Fano laser are $P_t = |A_t(t)|^2$ and $P_x = |A_x(t)|^2$, respectively.

Fig. S3 (b) shows the modulation principle of the Fano mirror. We modulate the second-order mode of the nanocavity (related to Fano mode) through index change which is caused by nonlinear effects (mainly band-filling and free-carrier dispersion (FCD), with free carriers excited by two-photon absorption (TPA)) [7] generated by the third-order mode of the nanocavity that is excited by external optical pulse [8]. The time-dependent complex frequency perturbations for the second- and the third-order modes are $\Delta\omega_{NL}$ and $\Delta\omega'_{NL}$, respectively, with the relationship of $\Delta\omega_{NL} = \eta' \Delta\omega'_{NL}$ ($\eta'$ is proportional to the ratio of their optical mode volumes). Here, $\Delta\omega'_{NL}$ can be expressed as [9]

$$\Delta\omega'_{NL} = -\left(K_{Kerr}|a'(t)|^2 - K_{Car}N_c(t)\right) - i\left(K_{TPA}|a'(t)|^2 + K_{FCA}N_c(t)\right), \qquad (s11)$$

where $K_{Kerr}$, $K_{Car}$, $K_{TPA}$, and $K_{FCA}$ are coefficients related to Kerr effect, carrier effects, TPA, and free carrier absorption (FCA), respectively [9]. $|a'(t)|^2$ is the energy of the third-order mode excited by the external modulating signal

$$\frac{da'(t)}{dt} = \left(-i\delta_c' - \gamma_t'\right)a'(t) + \sqrt{\gamma_v'}\eta_i s_m(t). \tag{s12}$$

Here $\delta_c' = \omega_0' + \Delta\omega_{NL}' - \omega_m$ is the detuning between the nanocavity resonance of the third-order mode $\omega_0' + \Delta\omega_{NL}'$ and the central frequency of the modulated signal $\omega_m$, where $\omega_0'$ is the unperturbed nanocavity resonance. Again, $\gamma_t' = \gamma_c' + \gamma_v'$, $\gamma_c'$, and $\gamma_v'$ are related to the corresponding Q-factors of the third-order mode by $\gamma_x' = \omega_0'/(2Q_x')$, $(x = v, c, t)$. $s_m(t)$ is the injected modulating signal and $\eta_i$ represents the injection efficiency. The carrier dynamics corresponding to the third-order mode can be described as

$$\frac{dN_c(t)}{dt} = \frac{\beta_{TPA}c^2}{2\hbar\omega_m n^2 V_{FCA}^2}|a'(t)|^4 - \frac{N_c(t)}{\tau_{recom}}, \tag{s13}$$

where $V_{FCA}$ is the effective FCA mode volume of the third-order mode, $\tau_{recom}$ is the free-carrier recombination time in the nanocavity. The typical parameters used in the calculations are: $Q_v = 1.2\times10^5$, $Q_c = 500$, $L = 5.527$ μm, $A_0 = 0.138$ μm$^{-2}$, $n = 3.17$, $\sigma = 10$, $\alpha_i = 160$ m$^{-1}$, $r_L = -0.955$, $\alpha = 3$, $\Gamma = 3.6\%$, $g_N = 9\times10^{-19}$ m$^2$, $N_{tr} = 1\times10^{23}$ m$^{-3}$, $\tau_s = 1$ ns, $\Delta\omega_{int} = 0.18\gamma_t$, $Q_v' = 8\times10^3$, $Q_c' = 4.5\times10^3$, $V_{TPA} = 0.484$ μm$^3$, $V_{FCA} = 0.378$ μm$^3$, $\tau_{recom} = 0.1$ ns, $\eta_i = 5.6\%$, $\eta' = 1$. These parameters are chosen based on ordinary semiconductor quantum well (QW) lasers and numerical simulations based on finite-difference time-domain (FDTD) and finite element methods.

As for the Fano laser, during cavity dumping, the lasing frequency $\omega_s$ changes due to the nanocavity resonance shift $\omega_0 \rightarrow \omega_0 + \Delta\omega$. We approximate the free carrier density change in the laser cavity by using the steady state solution as

$$\Delta N = -\frac{1}{\Gamma g_N}\frac{1}{\sigma L}\ln\left(\frac{|r_R(\omega_s)|}{|r_R(\omega_r)|}\right). \tag{s14}$$

Here $r_R(\omega_s) = \gamma_c/(i(\omega_0 + \Delta\omega - \omega_s) + \gamma_t)$ is the Fano mirror reflectivity at $\omega_s$, $\omega_r$ is the initial lasing frequency, and for simplicity, we set $\omega_r = \omega_0$. For a small perturbation ($\omega_0 + \Delta\omega - \omega_s \ll \gamma_t$), Eq. (s14) can be simplified to

$$\Delta N \approx \frac{1}{2\Gamma g_N \sigma L} \frac{(\omega_0 + \Delta\omega - \omega_s)^2}{\gamma_t^2}. \tag{s15}$$

At the same time, we can calculate the lasing frequency deviation $\omega_0 - \omega_s$ by using the phase $\phi_R(\omega)$ of the Fano mirror as

$$\phi_R(\omega_0) - \phi_R(\omega_s) + (2nL/c)(\omega_0 - \omega_s) + \sigma\Gamma\alpha g_N \Delta N L = 0. \tag{s16}$$

If $\alpha = 0$ for simplicity, we have

$$\omega_0 - \omega_s = -\frac{\Delta\omega}{1 + \gamma_t (2nL/c)}. \tag{s17}$$

The output power from the TP is $P_t(\omega_s) = |A_t(\omega_s)|^2 = |t_R|^2 |A^+(\omega_s)|^2$. In order to achieve a nanocavity resonance shift $\Delta\omega$ for the second-order (lasing) mode, the required nanocavity field change at the third-order mode $\Delta|a_c|^4$, can be estimated from the following equation

$$\Delta\omega = \eta' \Delta\omega' \approx \eta' K_{Car} G_{TPA} \tau_{recom} \Delta|a_c|^4 \approx \eta' K_{Car} G_{TPA} \tau_{recom} (\Delta E_c)^2, \tag{s18}$$

where $G_{TPA}$ accounts for the free carrier generation coefficient due to TPA [9]. Note that here, we only consider the band-filling and FCD-induced frequency shift. If the effective perturbation satisfies $\gamma_v \ll \omega_0 + \Delta\omega - \omega_s \ll \gamma_t$, which is the case in the experiment, we can derive analytical forms for the output power $P_t(\omega_s)$ and the contrast $ER_{FL}$ of the TP signal as

$$P_t(\omega_s) \approx A \left( \frac{R}{\sigma_s(\omega_s, N_s)} \right) \left( \frac{1}{B + \frac{C}{(\Delta E_c)^4}} \right)$$

$$ER_{FL} = \frac{P_t(\omega_s) - P_t(\omega_r)}{P_t(\omega_r)} = \frac{A}{D} \frac{1}{\sigma_s(\omega_s, N_s)} \left( \frac{1}{B + \frac{C}{(\Delta E_c)^4}} \right) - 1, \tag{s19}$$

where the parameters $A$, $B$, $C$, and $D$ are

$$\begin{aligned} A &= \Gamma V_p / (G_N \gamma_t^2) \\ B &= 1 / (2\Gamma g_N \sigma L \gamma_t^2) \\ C &= (N_r - N_{tr}) / \left( (1/(1 + c/(2nL\gamma_t)))^2 (\eta' K_{Car} G_{TPA} \tau_{recom})^2 \right) \\ D &= A\gamma_v^2 / (\sigma_s(\omega_r, N_r)(N_r - N_{tr})), \end{aligned} \tag{s20}$$

respectively, with $N_r$ being the threshold carrier density at the initial stage. $P_t(\omega_r)$ is the initial output power in the TP. The factor $\sigma_s$ is

$$\sigma_s = \frac{1}{\hbar\omega_s} \frac{(|r_L|+|r_{R,s}|)(1-|r_L||r_{R,s}|)}{2\gamma_{in}|r_L|\ln\{1/(|r_L||r_{R,s}|)\}}, \quad (s21)$$

in which the reflectivity amplitude of the Fano mirror is

$$|r_R(\omega_s)| = \gamma_c \bigg/ \sqrt{\gamma_t^2 + \left(1/\left(c/(2nL\gamma_t)+1\right)\right)^2 \left(\eta' K_{Car} G_{TPA} \tau_{recom}\right)^2 (\Delta E_c)^4}. \quad (s22)$$

Based on Eqs. (s19) – (s22), we can find that $P_t(\omega_s)$ increases monotonically with the pumping rate $R$ and the energy of the third-order mode $\Delta E_c$, while the contrast $ER_{FL}$ only increases monotonically with $\Delta E_c$. In addition, since $K_{Car}$ and $G_{TPA}$ are inversely proportional to the nonlinear optical mode volumes of the modulated mode (third-order mode), which usually scale with the nanocavity size and the linear optical mode volume (of both the second- and third-order modes), a smaller mode volume can lead to a smaller $C$, and thus a higher peak power and extinction ratio for a fixed $\Delta E_c$.

### A.3. Model of the gain-modulated Fabry–Pérot laser

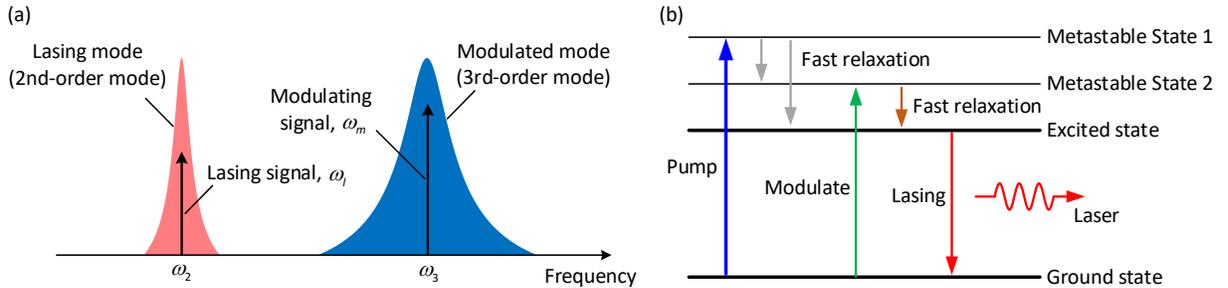

**Fig. S4**. Modulation principle of the FP laser. (a) The lasing frequency and the frequency of the modulating light are located at the second-order and third-order modes of the L7 nanocavity, respectively. (b) Four-level energy diagram for the FP laser.

The PhC line-defect laser can be viewed as a FP laser, which is lasing in the second-order mode (Fig. S4) but modulated via the third-order mode that is externally excited by optical pulses, similar to the Fano laser. The dynamics of the modulated mode can be described using the same form as Eq. (s12)

$$\frac{d}{dt}a_3(t) = \left(-i\delta_3 - \gamma_{t,3}\right)a_3(t) + \sqrt{\gamma_{v,3}}\eta_i s_m(t), \tag{s23}$$

except replacing $\Delta\omega_{NL}'$ with the complex nonlinear frequency change $\Delta\omega_3$ in $\delta_3 = \omega_3 + \Delta\omega_3 - \omega_m$, i.e., including the term related to the linear absorption due to the active material. Besides, the coupling decay rate between nanocavity and WG is absent here, but the absorption rate at transparency carrier density ($\gamma_{a,3} = v_g \alpha_{i,3}/2$) is included in $\gamma_{t,3} = \gamma_{v,3} + \gamma_{a,3}$, with $\alpha_{i,3}$ being the internal absorption loss at transparency carrier density. The complex nonlinear frequency change $\Delta\omega_3(t) = K_{D,3} N_3(t) + i K_{A,3}(N_{tr,3} - N_3(t))$ depends predominantly on the linear absorption with $K_{D,3} = \Gamma v_g \alpha_3 g_{N,3}/2$ and $K_{A,3} = -\Gamma v_g g_{N,3}/2$. All these parameters correspond to the values at the third-order mode frequency. Furthermore, $N_3(t)$ is the generated effective free carrier density at the modulated mode whose dynamics can be described as

$$\frac{d}{dt}N_3(t) = R_3 - \frac{N_3(t)}{\tau_s} + M_3\left(N_{tr,3} - N_3(t)\right)|a_3(t)|^2 - \frac{N_3(t)}{\tau_r}, \tag{s24}$$

where $R_3$ is the pumping rate at the modulated mode energy level, $M_3(N_{tr,3} - N_3(t))|a_3(t)|^2$ is the stimulated absorption rate, with $M_3 = (v_g g_{N,3})/(\hbar\omega_3 V_{p,3})$ and $V_3$ being the mode volume of the modulated mode. $N_3(t)/\tau_r$ represents the carrier relaxation from the level at the modulated mode frequency to the level at the lasing frequency, as enabled by the phonon and electron-electron scattering. For simplicity, we assume that such carrier coupling depends linearly on $N_3(t)$. $N_3(t)/\tau_s$ is the rest of the carrier decay rate. The effective free carrier density $N_2(t)$ at the lasing mode obeys the rate equation as

$$\frac{d}{dt}N_2(t) = R_2 - \frac{N_2(t)}{\tau_s} - M_2\left(N_2(t) - N_{tr,2}\right)|a_2(t)|^2 + \frac{N_3(t)}{\tau_r}, \tag{s25}$$

where $R_2$ is the pumping rate at the lasing mode energy level, $M_2(N(t) - N_{tr,2})|a_2(t)|^2$ is the stimulated emission rate, with $M_2 = (v_g g_{N,2})/(\hbar\omega_2 V_{p,2})$ and $V_2$ being the mode volume of the lasing mode. The dynamics of the lasing mode field have the same form as Eq. (s23), except that there is no input signal

$$\frac{d}{dt}a_2(t) = \left(-i\delta_2 - \gamma_{t,2}\right)a_2(t). \tag{s26}$$

The complex nonlinear frequency change $\Delta\omega_2$ in $\delta_2 = \omega_2 + \Delta\omega_2 - \omega_l$ is $\Delta\omega_2(t) = K_{D,2} N_2(t) + i K_{A,2}(N_{tr,2} - N_2(t))$ with $K_{D,2} = \Gamma v_g \alpha_2 g_{N,2}/2$ and $K_{A,2} = -\Gamma v_g g_{N,2}/2$.

The output power from the top of the nanocavity is $P_v(t) = \left|\sqrt{\gamma_{v,2}}\, a_2(t)\right|^2$. The typical parameters used in the calculations are: $Q_{v,2} = 1.5 \times 10^5$, $Q_{v,3} = 8 \times 10^3$, $g_{N,2} = 9 \times 10^{-19}$ m$^2$, $g_{N,3} = 9 \times 10^{-19}$ m$^2$, $\alpha_{i,2} = \alpha_{i,3} = 160$ m$^{-1}$, $N_{tr,2} = 0.4 \times 10^{23}$ m$^{-3}$, $N_{tr,3} = 0.41 \times 10^{23}$ m$^{-3}$, $\tau_s = 1$ ns, $\alpha_2 = 7$, $\alpha_3 = 0$, $V_{p,2} = 0.152$ μm$^3$, $V_{p,3} = 0.128$ μm$^3$, $\tau_r = 1.4$ ps.

From Eq. (s24), we have the nanocavity energy and effective free carrier density at steady state at the modulated mode as

$$\Delta E_{c,3} = |a_3|^2 = \frac{(1/\tau_s + 1/\tau_r) N_3 - R_3}{M_3 (N_{tr,3} - N_3)} . \tag{s27}$$

Since the pumping rate $R_3$ in the modulated mode is negligible, the carrier density in Eq. (s27) can be approximated as

$$N_3 \approx \frac{N_{tr,3}}{1 + \frac{(1/\tau_s + 1/\tau_r)}{M_3} \frac{1}{\Delta E_{c,3}}}, \tag{s28}$$

As seen, $N_3$ first increases with the nanocavity energy $\Delta E_{c,3}$ and then saturates ($N_3 \le N_{tr,3}$). The corresponding output power $P_v$ of the lasing mode can be derived as

$$P_v = \left|\sqrt{\gamma_{v,2}}\, a_2\right|^2 = \frac{\gamma_{v,2}}{M_2 (N_2 - N_{tr,2})} \left( R_2 - \frac{N_2}{\tau_s} + \frac{N_3}{\tau_r} \right). \tag{s29}$$

The contrast (the largest power change with respect to the initial power) can thus be calculated as

$$ER_{FP} = \frac{N_3}{\tau_r (R_2 - N_2/\tau_s)}. \tag{s30}$$

Based on Eqs. (s28) – (s30), we can find that the output power $P_v$ increases with the pumping rate $R_2$ and the energy of the modulated mode $\Delta E_{c,3}$, while the contrast $ER_{FP}$ decreases monotonically with the pumping rate $R_2$, which agrees with our measurements.

## B. Experimental setup

### B.1. Static measurement setup

The samples are vertically pumped by a 1480-nm continuous-wave (CW) laser using a micro-photoluminescence setup [10]. The Fano laser is pumped through a C-shape grating coupler [11] with a pumping efficiency of ~1%, while the FP laser is pumped from the top of the cavity. The pumping efficiency of the Fano laser is similar to that of the equivalent FP laser (PhC line-defect laser), which has a higher-order mode close to 1480 nm that facilitates absorption of the pump light. The full width at half maximum (diameter) of the pump spot is fixed at ~3 μm by an objective lens with a numerical aperture of 0.65. The emission is collected vertically using the same objective lens, and the collection efficiency ratio from the C-shape grating coupler with respect to the top of nanocavity is estimated to be 2 ~ 6 times. The lasing spectrum is measured at the 1550-nm port of the wavelength division multiplexer. All measurements are performed at room temperature. The pump power has been calibrated to the value after the objective lens.

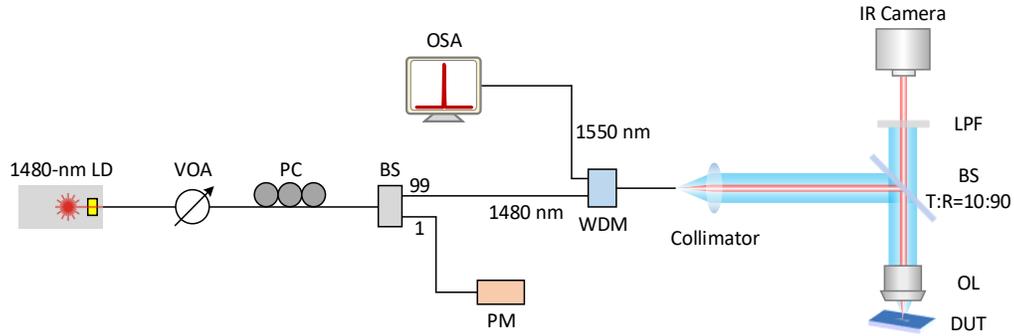

**Fig. S5.** Experimental setup for static measurement. LD: light diode; VOA: variable optical attenuator; PC: polarization controller; BS: beam splitter; PM: power meter; WDM: wavelength division multiplexers (1480 nm / 1550 nm); OSA: optical spectrum analyzer; OL: objective lens; LPF: long-wavelength pass filter; IR Camera: infrared camera; DUT: device under test. Black lines: standard single-mode fiber. Blue beam: 1480-nm CW light. Red beam: output lasing light.

**B.2. Dynamic measurement setup**

In addition to the 1480-nm CW pump path, another optical path is added for the modulating pulses (Fig. S6). The pulse is generated by a Ti:Sapphire laser cascaded with an optical parametric oscillator (INSPIRE HF100), and it has a pulsewidth of 170 fs with a repetition frequency of 79.9 MHz. The pulse first propagates through 10-m standard single-mode fiber to broaden the pulsewidth to ~ 4 ps so that it can be efficiently amplified by the following EDFAs. To obtain a sufficient modulating power, two-stage amplification is employed. Besides, since the modulating signal has to be aligned to the resonant mode of the L7 nanocavity, two optical band-pass filters are used to reshape its spectrum. The modulating light and pumping light are combined by a spatial beam splitter (T/R=50:50) before exciting the laser sample through the objective lens. The position and area of the pumping and modulating spots can be adjusted independently in order to optimize the coupling efficiencies. For the dynamic measurements of the Fano laser, the pumping and modulating signals are focused on the C-shape grating coupler and the side-coupled nanocavity, respectively, and the same channels are also used to collect the laser emissions, with the pumping (modulating) channel corresponding to the output signals from the TP (CP) of the Fano laser. For characterizing the dynamics of the FP laser, both the pumping and modulating signals are focused on the L7 nanocavity. The waveforms of the output signal are amplified by a low-power EDFA and filtered by a rectangle optical band-pass filter (bandwidth is 1.5 nm) before being measured by a communication signal analyzer (Agilent 86100A with module 86109B).

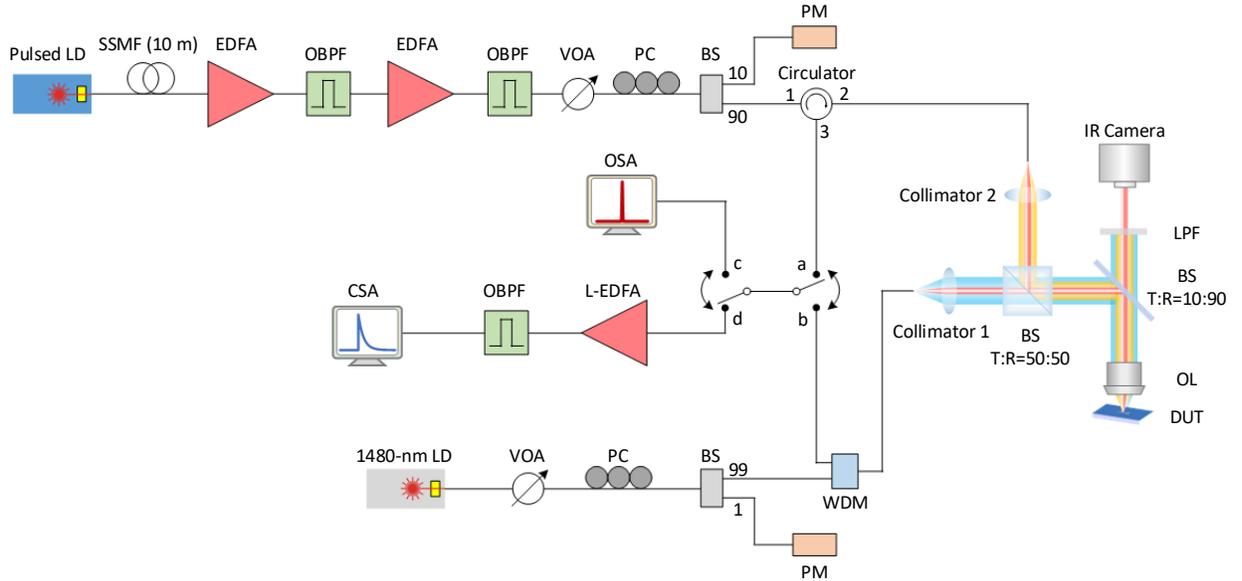

**Fig. S6.** Experimental setup for dynamic measurement. SSMF: standard single-mode fiber; (L-) EDFA: (low-power) Erbium-doped fiber amplifier; OBPF: optical band-pass filter; CSA: communication signal analyzer. Blue beam: input 1480-nm CW light; orange beam: input modulating signal; red beam: output lasing signal.

## C. Additional results

### C.1. Impact of the pump power and modulation power

According to the modulation mechanism of the Fano laser, the area of the blue component in the spectra reflects the generated pulse energy in the TP. It is shown that the area of the Fano laser keeps increasing with both the modulation power (Fig. S7 (a)) and the pump power (Fig. S7 (c)). In contrast, the blue spectral component of the FP laser, resulting from the increased free carrier density due to linear absorption of the modulation signal, saturates eventually (Fig. S7 (b) and Fig. S7 (d)). All of these are in accordance with their corresponding waveforms (see Fig. 4 in the main text). The evolution of the spectra of the Fano laser in dependence of the pump power (Fig. S7 (c)) differs from the case where the modulation power is varied (Fig. S7 (a)) because the amount of spectral broadening is mainly determined by the modulation power while the maximum intensity of the blue spectral component is dominated by the pump power. The spectral broadening of the FP laser saturates under high modulation power (Fig. S7 (b)) due to the carrier density saturation (Supplementary Note A.3). The spectral broadening is almost independent of the pump power (Fig. S7 (d)) because the gain is mainly provided by the linear absorption of the modulating pulse during the modulation process.

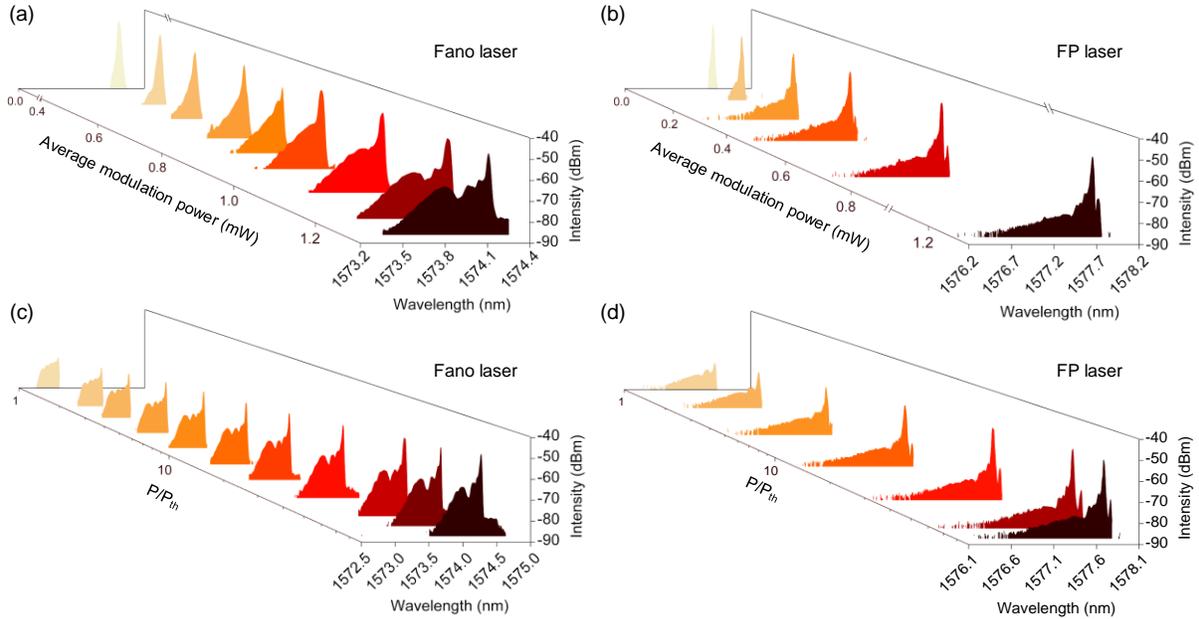

**Fig. S7**. (a) – (b) Measured spectra of the TP signal of the Fano laser (a) and the signal emitted from the top of the FP laser (b) for different values of the modulation power. The pumping rate is 63 $P_{th}$ for both lasers. The average modulation power refers to the value measured after the microscope's objective lens. (c) – (d) Measured spectra of the TP signal of the Fano laser (c) and the signal emitted from the top of the FP laser (d) with varied pump power. The average modulation power is fixed at 1.26 mW for both lasers.

## C.2. Simulated waveforms and contrast of the TP signal of the Fano laser

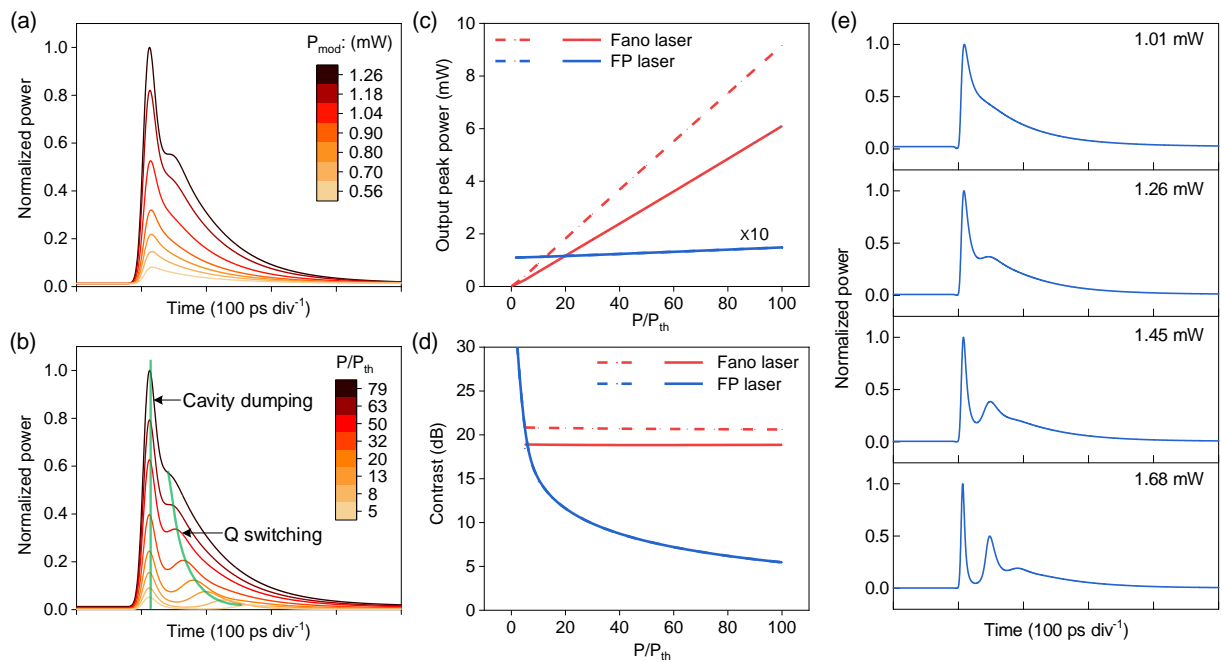

**Fig. S8**. (a) – (b) Simulated waveforms of the TP signal of the Fano laser for different modulation powers (a) and pump powers (b). The waveforms have been convolved with the impulse response (16.5 ps) of the oscilloscope, and normalized by the peak value of the highest pulse. In (b), the left green line indicates the peak associated with cavity dumping, and the right green curve indicates the peak associated with cavity Q-switching. (c) – (d) Comparison of the simulated output peak power (c) and signal contrast (d) versus pump power between the Fano laser (TP) and the FP laser. The solid (dashed) line is the calculated results after (before) the convolution with the oscilloscope's pulse response. The solid and dashed lines overlap in the FP laser. In (c), the output peak powers are amplified by 28 dB for comparison to experiments (due to the EDFA amplification in front of the oscilloscope), and the output peak power of the FP laser is multiplied by another factor of 10 to make it clearer for comparison. (e) Simulated (normalized) waveforms of the TP signal of the Fano laser for different modulation powers. The waveforms are not convolved with the impulse response of the oscilloscope. The pumping rate is fixed at 63 $P_{th}$ in (a) and (e), and the average modulation power is fixed at 1.26 mW in (b) – (d).

The simulated waveforms of the TP signal of the Fano laser (Fig. S8 (a) and (Fig. S8 (b)) agree well with the measurements (see Fig. 4 (a) and (d) in the main text). For the Fano laser, the output peak power starts from zero and increases linearly with the pump power (Fig. S8 (c)), deviating somehow from the sub-linearity of the measured curve (Fig. 4 (f) in the main text), which is mainly ascribed to two factors: one is the increased thermal effect under high pump power, which, for simplicity, has not been considered in the simulation, the other is the linear gain model used in the simulations. For the FP laser, the output peak power is finite when the pump power is absent, owing to the gain provided by the modulating pulse. The measured output power versus pump power (Fig. 4 (f) in the main text) decreases somehow at high pump powers, which is slightly different from the simulated results (Fig. S8 (c)), and we ascribe to the same factors as for the Fano laser. The signal contrast is constant for the Fano laser, while it decreases significantly for the FP laser as the pump power increases (Fig. S8 (d)). All of these are consistent with our analytical analysis in sections A.2 and A.3. Following the "primary pulse" generated due to cavity dumping, a "secondary pulse" is released due to cavity Q-switching (Fig. S8 (b)), in an agreement with the measured results (see Fig. 4 (d) in the main text). This process can also be well illustrated by the waveform evolution with different modulation power (Fig. S8 (e)). In addition, a narrower "primary pulse" can be achieved with a larger cavity perturbation, as implemented, e.g., by a higher modulation power.

### C.3. Dynamics of the Fabry–Pérot laser

On the arrival of the modulating pulse, free carriers are generated in the modulating mode (third-order mode) of the L7 nanocavity due to linear absorption, and then relax rapidly to the lasing mode (second-order mode), leading to gain modulation of the laser. Meanwhile, the lasing frequency also changes due to the linewidth-enhancement factor, resulting in an enhanced blue component in the lasing spectra (Fig. S9 (b)). The measurements can be well fitted by our theoretical model (Supplementary Note A.3), except for a major difference which is the dip in the leading edge of the measured waveform (Fig. S9 (a)). This is probably caused by the limitation of our linear gain model or other nonlinear absorption mechanisms, such as two-photon and free-carrier absorptions in the nanocavity, which are not included in our model.

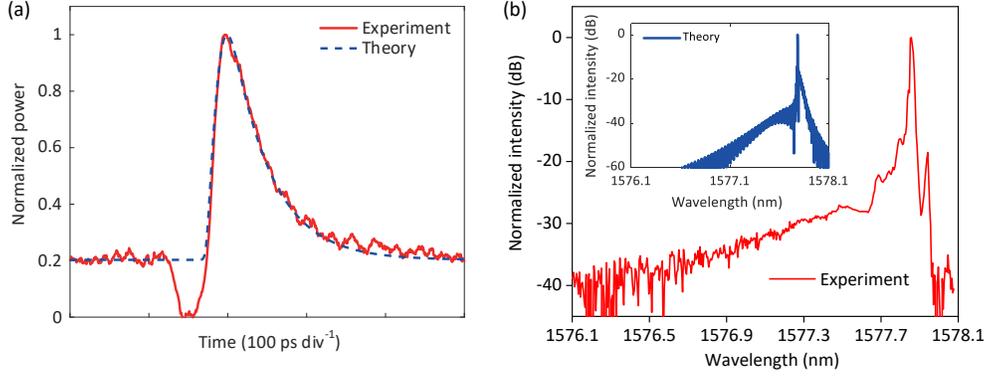

**Fig. S9**. Measured and calculated (normalized) waveforms (a) and their corresponding (normalized) spectra (b) of the FP laser. The pumping rate is 79 $P_{th}$, and the average modulation power is 1.26 mW.

### D. Impact of the EDFA noise

To accurately characterize the modulation contrast and output power of the lasers, it is essential to eliminate the impact of the EDFA noise on the measured waveforms. A commercial tunable CW laser at 1570 nm is amplified by our EDFA for the test. The relationships between the output power and noise background with the input power, as extracted from the spectrum, are characterized, cf. Fig. S10. As the laser power increases, the output power increases first and then saturates (Fig. S10 (a)). Meanwhile, the power of the noise background first decreases and then saturates (Fig. S10 (b)). This reflects the typical features of the EDFA, whose gain and spontaneous emission noise depend on the input power [12, 13]. The amplified signal can simply be described as

$$P_{out} = \frac{G_0}{1+\varepsilon_1 P_{in}} P_{in} + \frac{1}{1+\varepsilon_2 P_{in}} P_{sp}^0 . \tag{31}$$

Here, $P_{in}$ and $P_{out}$ are the input and output powers. $G_0$ is the linear gain, and $\varepsilon_1$ ($\varepsilon_2$) is the gain (noise) compression factor. By fitting the experimental results, we can obtain $G_0 = 95000$, $\varepsilon_1 = 1.2 \ \mu W^{-1}$, $\varepsilon_2 = 1.3 \ \mu W^{-1}$, $P_{sp}^0 = 78.7 \ \mu W$. For low input power $P_{in} < 0.1 \ \mu W$ (as in our experiments), $\varepsilon_1 P_{in} \ll 1$ and $\varepsilon_2 P_{in} \ll 1$, thereby, the output power of the signal and the noise background can be simplified as $P_{out} \approx G_0 P_{in} + P_{sp}^0$ and $P_{noise} \approx P_{sp}^0$. So the output power (noise background) depends linearly (does not depend) on the input power (see inset in Fig. S10 (a)). Under this condition, the output power $P'_{out}$ without the impact of the EDFA's noise can be subtracted as $P'_{out} = P_{out} - P_{noise}$.

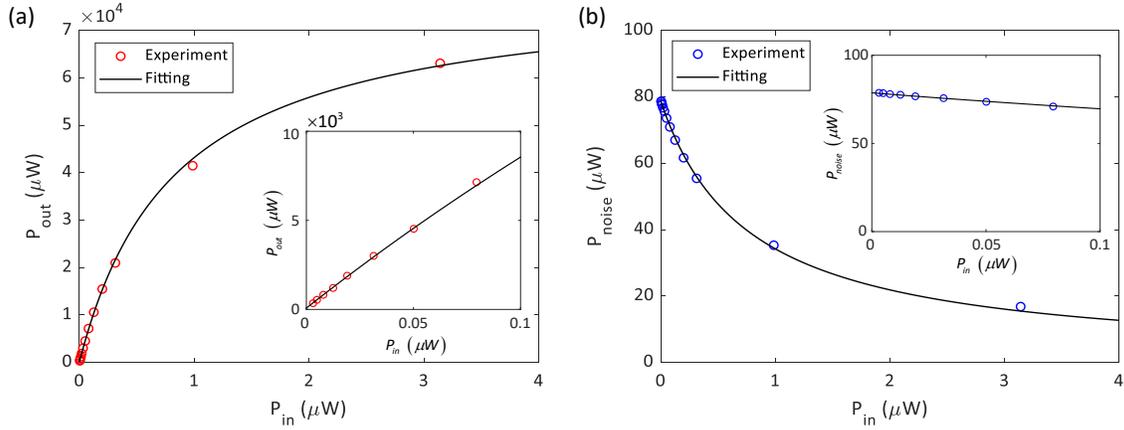

**Fig. S10**. Measured output peak power (a) and noise power (b) of the optical spectra after EDFA amplification versus the input power. Inset in (a) and (b): zoom-in of (a) and (b) at low input power, respectively.